\documentclass[3p,twocolumn]{elsarticle}

\usepackage{lineno,hyperref}
\modulolinenumbers[5]

\journal{Journal of \LaTeX\ Templates}

\biboptions{authoryear}

\usepackage{color}
\usepackage{amssymb}

\def\hi{H\,{\sc i}}

\def\msun{M$_{\odot}$}

\begin{document}

\begin{frontmatter}

\title{Measurements of the angular momentum-mass relations in the \textsc{Simba} simulation}

\author[UWC]{E. Elson}
\author[UWC,IDIA,ICRAR]{M. Glowacki}
\author[UWC,UEd,SAAO]{R. Dav\'{e}}

\address[UWC]{Department of Physics $\&$ Astronomy, University of the Western Cape,\\ Robert Sobukwe Rd, Bellville, 7535, South Africa.}
\address[ICRAR]{International Centre for Radio Astronomy Research, Curtin University, Bentley,\\ WA 6102, Australia}
\address[IDIA]{Inter-University Institute for Data Intensive Astronomy, Bellville 7535, South Africa}
\address[UEd]{Institute for Astronomy, Royal Observatory, University of Edinburgh,\\ Edinburgh EH9 3HJ, UK}
\address[SAAO]{South African Astronomical Observatory, Observatory, Cape Town 7925, South Africa}

\begin{abstract}
We present measurements of the specific angular momentum content ($j$)  of galaxies drawn from the \textsc{Simba} cosmological hydrodynamic simulations.  For the stellar, \hi\ and baryonic matter components we demonstrate the existence of extremely tight relations between $j$ and the mass contained within the radius at which the \hi\ mass surface density decreases to 1~\msun~pc$^{-2}$.  These relations are broadly consistent with a variety of empirical measurements.  We confirm the observational result that the scatter in the stellar $j$--$M$ relation is driven largely by \hi\ content, and measure the dependence of its scatter on the deviations of galaxies from other important scaling relations. For a given stellar mass,  \hi-rich/poor galaxies have more/less-than-average stellar specific angular momentum.  A similar, yet weaker, correlation exists for \hi\ mass fraction.  Overall, our results demonstrate the utility of the \textsc{Simba} simulations as a platform for understanding and contextualising the data and results from forthcoming large galaxy surveys. 
\end{abstract}

\end{frontmatter}

\section{Introduction}
An important goal in modern astronomy is to quantitatively understand the processes by which galaxies build up their mass in a hierarchical formation scenario.  The dynamical characteristics of a galaxy serve as a proxy for the formation history of the system, including interactions with other galaxies.   In a $\Lambda$CDM Universe, the growth of structure is governed by the tidal field of an irregular matter distribution, resulting in the transfer of angular momentum from the tidal field to collapsing proto-galaxies (e.g., \citealt{Peebles_1969}).  Angular momentum therefore plays a key role in the formation and evolution of galaxies. 

In recent years, several investigators have carried out  measurements of the specific angular momentum content of nearby observed galaxies.  The manner in which the measurement is made depends to a large extent on the characteristics of the available data.  The simplest approach that is often applied to relatively large samples of galaxies involves approximating $j$ by adopting a constant-value measure of a galaxy's rotation velocity (such as $W_{50}/2\sin i$) and a particular measure of its radial extent (such as effective radius).  While the method is relatively quick and easy to apply, it can lead to the introduction of random and systematic errors in $j$, which need to be reliably estimated.  \citet{Fall_1983} used this approach to demonstrate the existence of a power-law relation between the stellar specific angular momenta, $j_*$, of a sample of galaxies and their stellar masses, $M_*$.  \citet{Romanowsky_2012} extended the study of \citet{Fall_1983} by carrying out what was the largest investigation of galaxies in the $j_*-M_*$ plane to date.  More precise measurements of $j$ can be generated for galaxies that are both spatially and kinematically well-resolved.  Kinematic maps from \hi\ interferometers and optical or infrared telescopes equipped with an integral field spectroscopy unit are well suited to producing high-precision observational measurements of $j$. The pioneering work of \citet{Obreschkow_2014} focused on using azimuthally-averaged velocity and mass-density profiles of a sample of 16 spiral galaxies from The \hi\ Nearby Galaxy Survey \citep{THINGS_2008} to measure $j$ as a function of galactocentric radius.  The scatter in their relations was shown to be driven largely by bugle mass fraction. \citet{cortese_2016} used 2D resolved line-of-sight velocity maps of the stellar and ionised gas content of 488 galaxies from the SAMI survey \citep{Croom_2012} to show that $j_*$ increases monotonically with $M_*$, while the scatter in the relation is strongly correlated with optical morphology.  \citet{Posti_2018} applied the high-precision method to 92 galaxies from the \textit{Spitzer} Photometry and Accurate Rotation Curves sample \citep{SPARC} to produce an improved estimate of the $j_*$--$M_*$ relation over a large range of stellar mass.  More recently, \citet{Hardwick_2022} used a hybrid approach based on single-value measures of velocity (from single-dish \hi\ velocity width measurements) and measures of the spatially-resolved mass content of 564 nearby galaxies in the eXtended GALEX Arecibo SDSS Survey \citep{xGASS} to show the scatter in the local $j_*$--$M_*$ relation to be driven mainly by \hi\ gas fraction (at lower masses) and bulge fraction (at higher masses).

Various galaxy properties have been identified as contributors to scatter in measured $j$--$M$ relations.  Galaxy morphology is a main driver.  The study of \citet{Romanowsky_2012} used extended kinematic data for $\sim 100$ nearby bright galaxies of all types.  They showed ellipticals and spirals to form two parallel tracks in the $j_*$--$M_*$ plane.  The improved accuracy of measurements offered by the study of \citet{Obreschkow_2014} based on spatially-resolved multi-wavelength imaging led to the discovery of galaxy bulge mass fraction being the dominant driver in the baryonic $j$--$M$ relation.  \citet{cortese_2016} showed  that for stellar masses $> 10^{9.5}$~\msun, the scatter in the $j_*$--$M_*$ relation is related to the stellar light distribution, hence morphology, of galaxies. They attribute this finding to the fact that, at fixed stellar mass, the contribution of ordered motions to the dynamical support of galaxies varies significantly.  \citet{Posti_2018} measured that the Fall relation extends to dwarf galaxies as a single, unbroken power-law. The showed that the tightness of the relation is improved if one considers only the disk components of spiral galaxies, and that its slope is closer to that of dark matter haloes. \citet{Gillman_2019} used adaptive optics integral field observations of 34 star-forming galaxies  from $0.08 \le z \le 3.3$ to measure inclination-corrected rotational velocities, half-light sizes and stellar masses, and investigated how the $j_*$--$M_*$ relation for their sample evolves with cosmic time.  They demonstrated an evolution in angular momentum content that is likely driven by the internal redistribution of angular momentum from the accretion of material.  \citet{Pina_2021} measured the stellar and gas specific angular momenta for a sample of nearby galaxies, and discussed their internal correlations.  They found disk scale length and gas fraction to be significant contributors to scatter in their relations.  \citet{Pina_2021b} extended the efforts of \citet{Pina_2021}, further showing gas-rich galaxies to have larger stellar and baryonic momenta (but lower gas momenta) than gas-poor galaxies, at fixed mass.

Several authors have used cosmological simulations to study the $j$--$M$ relation.  \citet{genel_2015} used the Illustris cosmological simulation \citep{Illustris} to generate distinct stellar $j$--$M$ relations for late-type and early-type galaxies.  They showed the relations to correspond to different retention factors of the specific angular momentum generated by cosmological tidal torques. They further found feedback (in the forms of galactic winds and active galactic nuclei) to play a role in controlling the relations.  \citet{Teklu_2015} also studied the stellar $j$--$M$ relation using hydrodynamical cosmological simulations taken from the set of Magneticum Pathfinder simulations\footnote{www.MAGNETICUM.org}.  Among other things, they showed the specific angular momentum of the stars in disk galaxies is slightly smaller compared to that of the cold gas.  \citet{Obreja_2016} showed galaxies from  simulations taken from the  Numerical Investigation of a Hundred Astrophysical Objects (NIHAO) project \citep{NIHAO} to fall on top of observed ones in the stellar $j$--$M$ plane.   The disc and spheroid components of the galaxies show shallower correlations between $j_*$ and $M_*$ than the theoretical expectations for pure spirals and ellipticals.  \citet{Lagos_2017} used the {Evolution and Assembly of GaLaxies and their Environments simulation suite \citep{EAGLE} to study the redshift evolution of specific angular momentum within galaxies.  They found $j_*$ and $j_\mathrm{b}$ (baryonic angular momentum) to be strongly correlated with stellar and baryonic mass, and showed the scatter in the relations to correlate with various morphological proxies such as gas fraction, stellar age, etc.  \citet{Stevens_2018a} used the Dark Sage semi-analytic model \citep{DarkSAGE} to study the connection between atomic gas fraction and a global instability parameter determined by the ratio of disk specific angular momentum to mass.  \citet{Stevens_2018a} demonstrated a clear correlation between gas fraction and disk specific angular momentum  at fixed disk mass.  Semi-analytic models were also used by \citet{Zoldan_2018} to predict a strong dependence of  the specific angular momentum of galaxies on their cold gas content. Spirals decomposed into bulge and disk subcomponents were found  to follow similar trends.  \citet{El-Badry_2018} used a suite of cosmological zoom-in simulations from the Feedback In Realistic Environments project \citep{FIRE} to show that the ratio of specific angular momentum of gas in the central galaxy to that of its dark matter halo increases significantly with galaxy mass.  They claim the reason for the reduced angular momentum of baryons in low-mass haloes to be due to the fact that low-mass haloes accrete gas less efficiently at late times, when the mean specific angular momentum content of accreted gas is highest. 

In this work, we use the \textsc{Simba} cosmological hydrodynamical simulations \citep{Simba_2019} to measure relationships between specific angular momentum and mass for various mass components of galaxies.  We compare our results to those based on observational data, and use the information that \textsc{Simba} provides  to search for  systematic variations of scatter in the stellar $j$--$M$ relation with scatter in other important scaling relations.  Demonstrating that the \textsc{Simba} simulations  produce $j$--$M$ relations that are consistent with empirical results serves to further promote them as a platform for understanding and interpreting data and results from forthcoming galaxy surveys. 

The layout of this paper is as follows.  We present in Section~\ref{section_data} the sample of \textsc{Simba} galaxies we use for this study, as well as our method of measuring their specific angular momentum content.  In Section~\ref{section_results} we show the $j$--$M$ relations we measure for the various mass components of the \textsc{Simba} galaxies.  We offer a discussion of the results in Section~\ref{section_discussion}, and investigate the properties of the residuals in Section~\ref{residuals}.  Finally, our results are summarised in Section~\ref{section_summary}.


\section{Data}\label{section_data}
\subsection{Simulations}
In this work we used the \textsc{Simba} cosmological galaxy formation simulations \citep{Simba_2019} to measure the \hi\ dynamics of galaxies. \textsc{Simba} is run with \textsc{Gizmo's} meshless finite mass hydrodynamics and includes detailed prescriptions for sub-resolution star formation and feedback processes, including black hole growth. We utilised the high-resolution box consisting of $512^3$ dark matter particles and $512^2$ gas elements within a $(25~h^{-1}~\mathrm{Mpc})^3$ periodic volume.  The resulting mass resolutions for dark matter particles and gas elements are $1.2\times 10^7$~\msun\ and $2.28\times 10^6$~\msun, respectively. 

\textsc{Simba} does not explicitly model (owing to resolution limitations) physical processes giving rise to the cold phase of the interstellar medium.  Self-shielding and a  prescription for H$_2$ fraction based on gas metallicity work together to transform ionised gas into atomic and molecular phases.   The chemical enrichment model used by \textsc{Simba} tracks 9 metals during the simulation, tracking enrichment due to Type Ia and Type II supernovae and asymptotic giant brach stars.  Star formation is modelled using an H$_2$-based \citet{Schmidt_1959} relation that uses the H$_2$ density and dynamical time.  A \citet{Chabrier_2003} stellar initial mass function is assumed in order to compute stellar evolution.  \textsc{Simba} uses a 6-D friends-of-friends algorithm to group star and gas particles into galaxies.  The interested reader is referred to \citet{Simba_2019, Dave_2020} and \citet{Glowacki_2020} for further details regarding star formation, feedback and other galaxy formation physics employed by \textsc{Simba}.

\subsection{Sample}
We used a sample of \textsc{Simba} galaxies taken from \citet{Glowacki_2021}.  Those authors applied the following \hi\ mass, stellar mass and specific star formation rate cuts to the galaxies in the high-resolution \textsc{Simba} snapshot: $M_\mathrm{HI}>1.25\times 10^8$~\msun, $M_*>7.25\times 10^8$~\msun, $\mathrm{sSFR}>10^{-11}$~yr.  These cuts were made in order to build a sample of galaxies that are actively forming stars and which have masses above the numerical resolution limits of the high-resolution  simulation.  An additional cut was made on dynamical morphology in order to ensure the galaxies have their dynamics dominated by rotation.  Galaxies with $\kappa_\mathrm{rot}>0.7$ were selected as being disk-dominated, where $\kappa_\mathrm{rot}$ is the fraction of energy invested in ordered rotation \citep{Sales_2012}.  Finally, \hi\ total intensity maps of all galaxies surviving these cuts were generated and visually inspected in order to further remove a small number of galaxies that have experienced a recent merger. The final galaxy sample is of size 187. We treat this sample as being representative of nearby spiral galaxies that are gas-rich, star-forming, and that have well-defined disks dominated by rotation.   Distributions of the stellar masses, \hi\ masses and baryonic masses for the galaxies are shown in Figure~\ref{distribs1}.

\begin{figure*}
\centering
	\includegraphics[width=1.5\columnwidth]{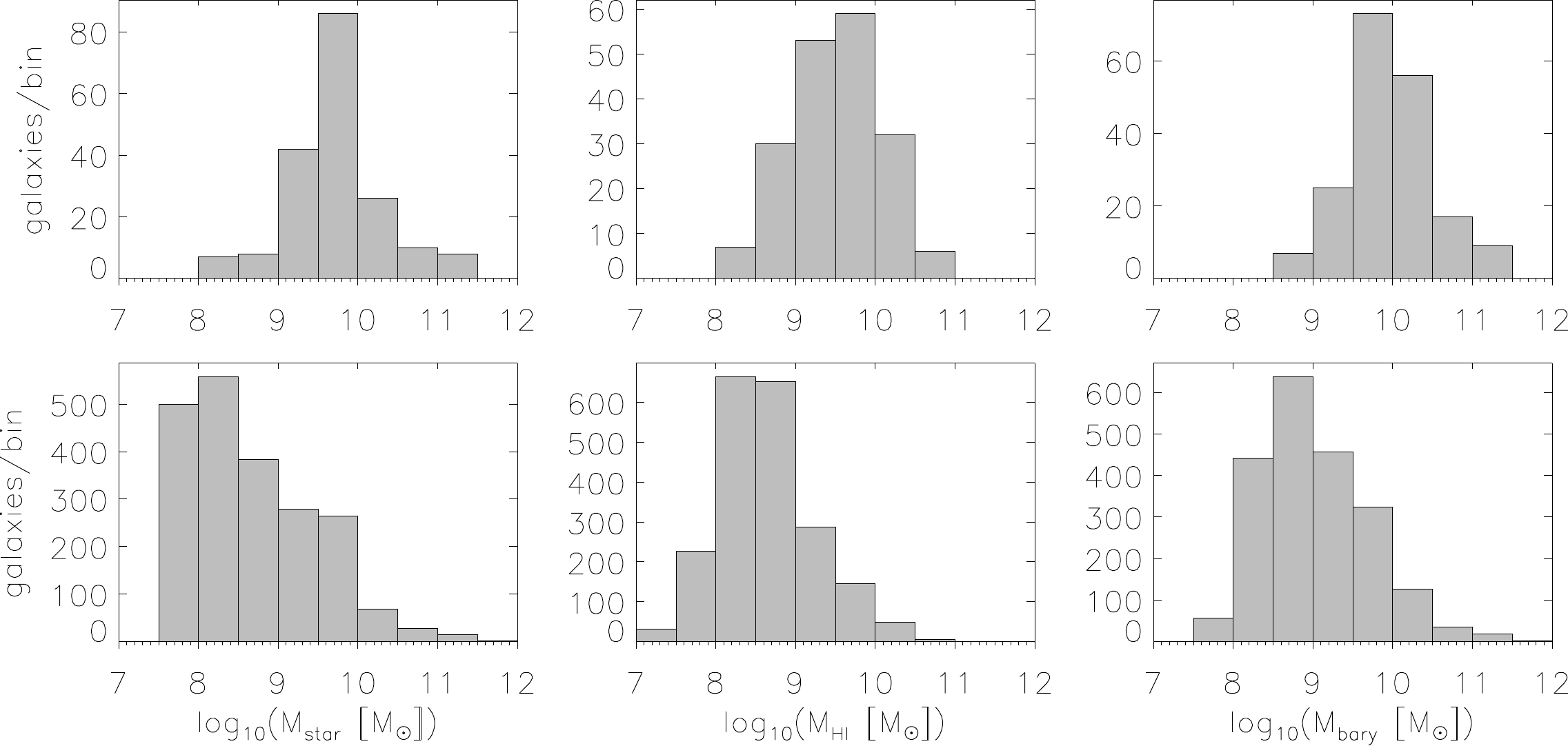}
    \caption{Top row: Distributions of stellar mass, \hi\ mass and baryonic mass for the 187 galaxies taken from \citet{Glowacki_2021}.  Bottom row: For comparison, the same distributions shown for all galaxies in the \textsc{Simba} snapshot from which the \citet{Glowacki_2021} sub-sample of 187 galaxies was selected.}
    \label{distribs1}
\end{figure*}

\subsection{Method}
Given a simulated galaxy consisting of $N$ particles, $\vec{j}$ is calculated as
\begin{equation}
\vec{j}={\sum_i m_i(\vec{r}_i-\vec{r}_\mathrm{COM})\times(\vec{v}_i-\vec{v}_\mathrm{COM})\over \sum_i m_i}, 
\label{j_eqn_particles}
\end{equation}
where $\vec{r}_i$ and $\vec{v}_i$ are the position and velocity vectors of particle $i$, $\vec{r}_\mathrm{COM}$ and $\vec{v}_\mathrm{COM}$ are the position and velocity vectors of the centre of mass (of all the particles), and $m_i$ is the mass of particle $i$.  This calculation is carried out for each of a galaxy's stellar, \hi\ and baryonic mass components.   In order to ensure an accurate calculation of a galaxy's \hi\ angular momentum content, we considered only the fraction of each gas particle's mass that corresponds to neutral atomic hydrogen.

In order to make our results more directly comparable to empirical results in the literature, we consider only the particles that are within a radius $R_\mathrm{HI}$ of their centre of mass, where $R_\mathrm{HI}$ is the radius at which the \hi\  mass surface density of a galaxy drops to 1~\msun~pc$^{-2}$.  This is the radius typically used by observers to mark the \hi\ edge of a galaxy.  We did also carry out all analyses using radius limits of 2$R_\mathrm{HI}$ and 3$R_\mathrm{HI}$ and found there to be no significant changes in our main results.  

\section{Results}\label{section_results}
Figure~\ref{j-M_profiles1} shows in its left column $\log_{10}j$ as a function of $\log_{10}M$, with both quantities  measured at  radius $R=R_\mathrm{HI}$ for the stellar, \hi\ and baryonic mass components of the \textsc{Simba} galaxies.  All $j$ measurements have been made using Equation~\ref{j_eqn_particles}.  For each $j$--$M$ relation, we carried out a least-squares fit to $\log_{10}j$ as a function of $\log_{10} M$:
\begin{equation}
\log_{10} j = \alpha(\log_{10}(M/M_{\odot})-11)+\beta.
\label{jmodel}
\end{equation}
This is equivalent to a power-law dependence ($j\propto M^{\alpha}$).  In order to generate uncertainty measures for $\alpha$ and $\beta$, we carried out 100 fits to each $j$--$M$ relation, each time using a 50\% random subset of the data.  The uncertainties we quote are the differences between the 90-th and 10-th percentiles of the resulting parameter sets.  Each panel in the right column of Figure~\ref{j-M_profiles1} shows the distribution of the vertical distances between the data points and the best-fit model.  The parameters and statistics for the various $j$--$M$ relations are collected in Table~\ref{j-M_table}.  

\begin{figure*}
\centering
\includegraphics[width=1.5\columnwidth]{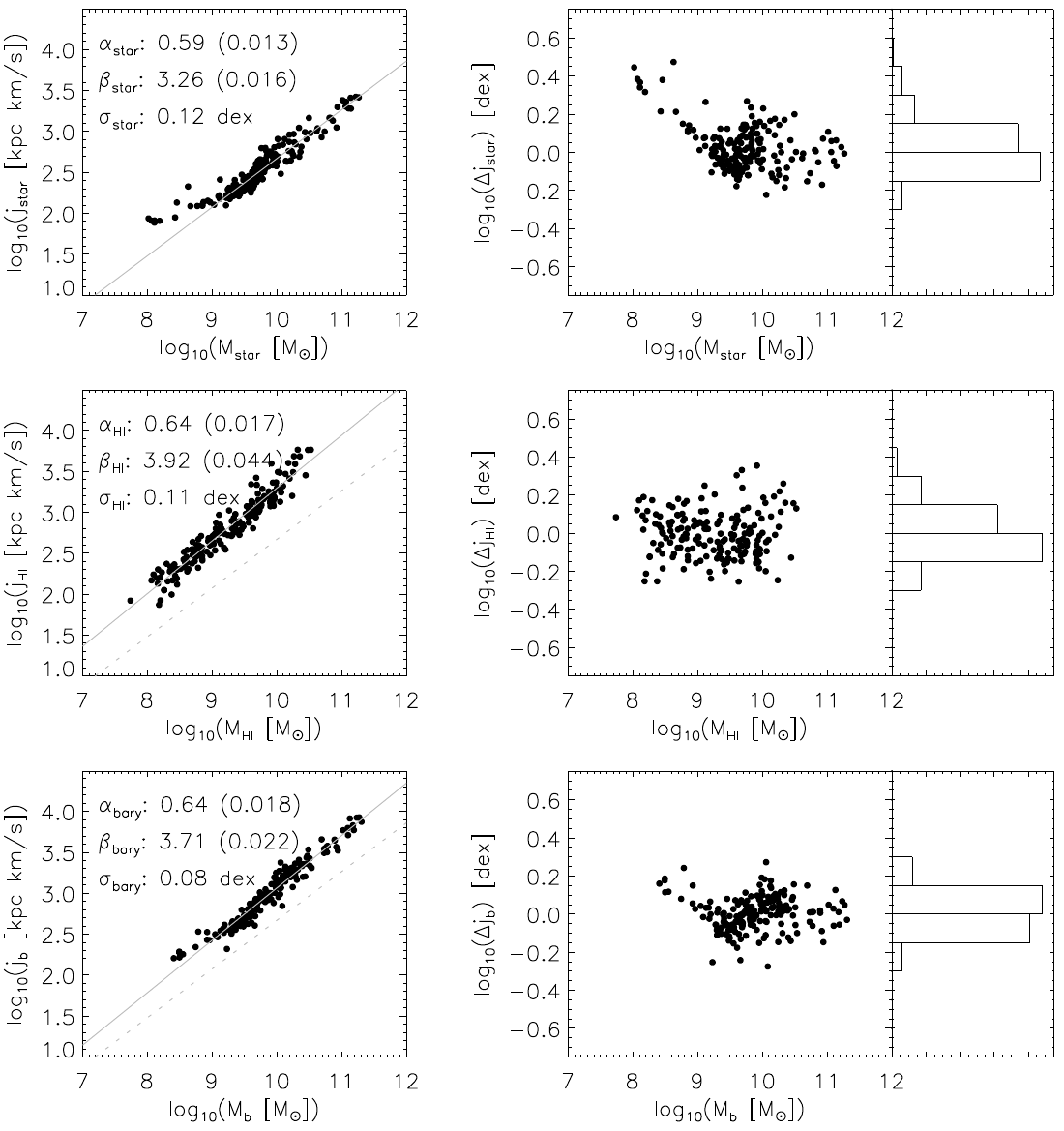}
\caption{Left column: $j$--$M$ relations for the stellar, \hi\ and baryonic mass components of \textsc{Simba} galaxies.  The solid line in each panel is the result of a least-squares fit of the model $\log_{10} j = \alpha(\log_{10}(M/M_{\odot})-11)+\beta$ to the data.  The best-fit $\alpha$ and $\beta$ values are shown in the top left of each panel.  The best-fit line for the stellar $j$--$M$ relation is reproduced as a dotted line in the panels for the other $j$-$M$ relations, in order to facilitate comparisons.  Middle column: Residuals measured as vertical separations between the data points and the best-fit relation, shown as a function of stellar mass. Right column: Distribution of the aforementioned residuals.}
\label{j-M_profiles1}
\end{figure*}

\begin{table}
{\footnotesize
\caption{Best-fit parameters ($\alpha$ and $\beta$) for the $j$--$M$ relations presented in Figure~\ref{j-M_profiles1}, as well as the standard deviation of the vertical distances of the points about each fitted relation ($\sigma_{\perp}$).}
\begin{center}
\begin{tabular}{llll}
\hline
Mass component	&	$\alpha$	&	$\beta$	&	$\sigma_{\perp}$ 	\\
\hline
stars				&	$0.59\pm 0.013$		&	$3.26\pm 0.016$		&	0.08				\\
\hi\				&	$0.64\pm 0.017$		&	$3.92\pm 0.044$		&	0.11				\\
baryons			&	$0.64\pm 0.018$		&	$3.71\pm 0.022$		&	0.08				\\
\hline
\end{tabular}
\end{center}
\label{j-M_table}}
\end{table}%

The $j$--$M$ relations we measure for the \textsc{Simba} galaxies are all clearly linear (in log-log space) with  small amounts of  scatter.  The \hi\ relation is consistent with a single, unbroken power law that describes both dwarf and more massive galaxies.  The same is true for the stellar and baryonic relations for masses $\gtrsim 10^9$~\msun.  Below this mass limit there is a clear upturn in both of the relations.  This is likely due to the fact that in the \textsc{Simba} simulations, galaxies with $\log M_*< 9.5$~\msun\ do not experience any black hole feedback. When black hole feedback does turn on at $\log M_*=9.5$~\msun, it stalls $M_*$ growth, which causes a pile up of galaxies below this mass threshold.  Our power-law fits to the stellar and baryonic relations consider only those galaxies above $10^9$~\msun.  We discuss the relations in further detail in the next section.

\section{Discussion}\label{section_discussion}
In this section, we discuss our parameterised \textsc{Simba} $j$--$M$ relations and compare them to empirical results from the literature. 

Traditionally, the stellar relation has drawn the most attention.  The  parameters of $\alpha =0.59 \pm 0.013$ and $\beta = 3.26\pm 0.016$  measured for our $j_*$--$M_*$ relation compare favourably with a variety of results from the literature, some of which we list here.  \citet{Romanowsky_2012} used approximations for the specific angular momentum content of spiral and elliptical galaxies to place $\sim~100$ nearby bright galaxies on a $j_*$--$M_*$ diagram.  For their entire sub-sample of spirals, they measured $\alpha=0.61\pm 0.04$ and $\beta=3.31\pm 0.02$ with a residual rms scatter of 0.20 dex.  They found similar trends for the bulge and disk subcomponents of their spiral galaxies.  \citet{cortese_2016} used homogenous resolved velocity maps for a large sample of galaxies to show that at fixed stellar mass, disk-dominated systems have higher specific angular momentum than bulge-dominated galaxies.  For their full sample, their power-law index is measured to be $0.64 \pm 0.04$.  \citet{Posti_2018} accurately determined the empirical $j_*$--$M_*$ relation for 92 nearby galaxies spanning morphologies from S0 to Irr and the stellar mass range $7\leq \log M_*/M_\mathrm{\odot}\leq 11.5$.  Over this entire mass range, they found the relation to be well-described by a single power-law with index $\alpha=0.55\pm 0.02$ and with intercept and $\beta=3.34\pm 0.03$.  \citet{Pina_2021} used a larger sample of nearby galaxies (from a variety of surveys) to generate an almost identical result: $\alpha=0.54\pm 0.02$.}

While the stellar $j$--$M$ relation was first studied as far back as 1983, the gas and baryonic relations have received attention much more recently.  The pioneering study of \citet{Obreschkow_2014} offered high-precision measurements of the specific angular momenta contained in the baryonic components of 16 nearby spiral galaxies from the THINGS sample.  They found $M$, $j$ and $\beta$ (bulge fraction) to be strongly and irreducibly correlated.  Late-type galaxies were shown to scatter around a mean relation $j\propto M^{2/3}$ while any subsample of fixed $\beta$ followed a relation $j\propto M$.  \citet{Pina_2021} also studied the baryonic $j$--$M$ relation, finding it to be well-modelled by a single power-law with slope $\alpha=0.60\pm 0.02$ over a large mass range that includes dwarf galaxies.  Other authors had suggested that dwarf galaxies have higher-than-expected fractions of  baryonic specific angular momentum, and invoked scenarios involving feedback processes and cold gas accretion to explain their findings.  However,  results such as ours and those of \citet{Pina_2021} - results that indicate the existence of a single power-law that well-describes the full mass range of galaxies - suggest that feedback and accretion processes work in similar ways for most galaxies.

Several studies have focused on the gaseous $j$--$M$ relation in recent years. \citet{Butler_2017} extended the efforts of \citet{Obreschkow_2014} to 14 low-mass, gas-rich galaxies from the LITTLE THINGS sample.  They found the specific angular momentum of \hi\ to be about 2.5 times higher than that of the stars. \citet{Pina_2021} again found an unbroken power-law for the $j_\mathrm{gas}$--$M_\mathrm{gas}$ relation based on their galaxy sample, yet one with a higher slope of $\alpha=1.02\pm 0.04$.   They again emphasise the fact that dwarfs seems to obey the same relation as high-mass galaxies.  In this work, we do not find the \hi\ $j$--$M$ relation to have a slope that is higher than those of the stellar and baryonic relations. Rather, we find the slopes of the \hi\ and the baryonic $j$--$M$ relations in \textsc{Simba} to be essentially identical, and very close to 2/3.  The zero-point of our \hi\ $j$--$M$ relation is a factor $\sim 4.6$ higher than that of the stellar relation, which is similar, yet more extreme, than the finding of \citet{Butler_2017}.  

In this study, we have not considered the dark matter $j$--$M$ relation of the \textsc{Simba} galaxies.  Theoretical arguments are often used to show $j_\mathrm{h}\propto M_\mathrm{h}^{2/3}$ for dark matter haloes (e.g., see \citealt{Obreschkow_2014}).  However, our finding that the stellar, \hi\ and baryonic $j$--$M$ relations for the \textsc{Simba} galaxies have power-law indices close to 2/3 do not necessarily imply the same value for dark matter haloes.  Some authors (e.g., \citealt{Liao_2017}) have shown $j_\mathrm{h}\propto M_\mathrm{h}^{2/3}$ to be the case for particular numerical simulations, yet further investigation is needed.

\section{Residuals}\label{residuals}
Several authors have investigated the possibility that the scatter in the $j_*$--$M_*$ relation varies systematically with other galaxy parameters.  \citet{Fall_1983} suggested that the Hubble sequence may be related to the systematic variation in $j_*$ at a fixed $M_*$.  Using a larger sample of galaxies spanning a broader morphology range, \citet{Romanowsky_2012} showed that the scatter correlates with morphology across the Hubble sequence.  The high-precision measurements of \citet{Obreschkow_2014} showed bulge mass fraction to be strongly correlated with specific baryon angular momentum and mass. \citet{cortese_2016} confirmed these findings for a larger sample, showing both visual morphology and Sersic index within one effective radius to contribute to $j_*$--$M_*$ scatter. \citet{Posti_2018} showed that the stellar $j$--$M$ relation becomes tighter when considering only the disc components of spiral galaxies (i.e. by removing the bulge contribution to the light profile).  The result, they say, possibly indicates a simpler link to dark matter haloes of discs compared to bulges.  \citet{Pina_2021} found internal correlations inside their various $j$--$M$ relations, showing disc scale length and gas richness to be significant drivers of scatter. \citet{Pina_2021b} demonstrated tight correlations between $j$, $M$ and the cold gas fraction of the interstellar medium.  An important implication of these findings is that, statistically, important galaxy properties such as morphology (or other structural parameters) can be predicted once the mass and specific angular momentum of the galaxy are known.

\begin{figure*}
\centering
\includegraphics[width=1.5\columnwidth]{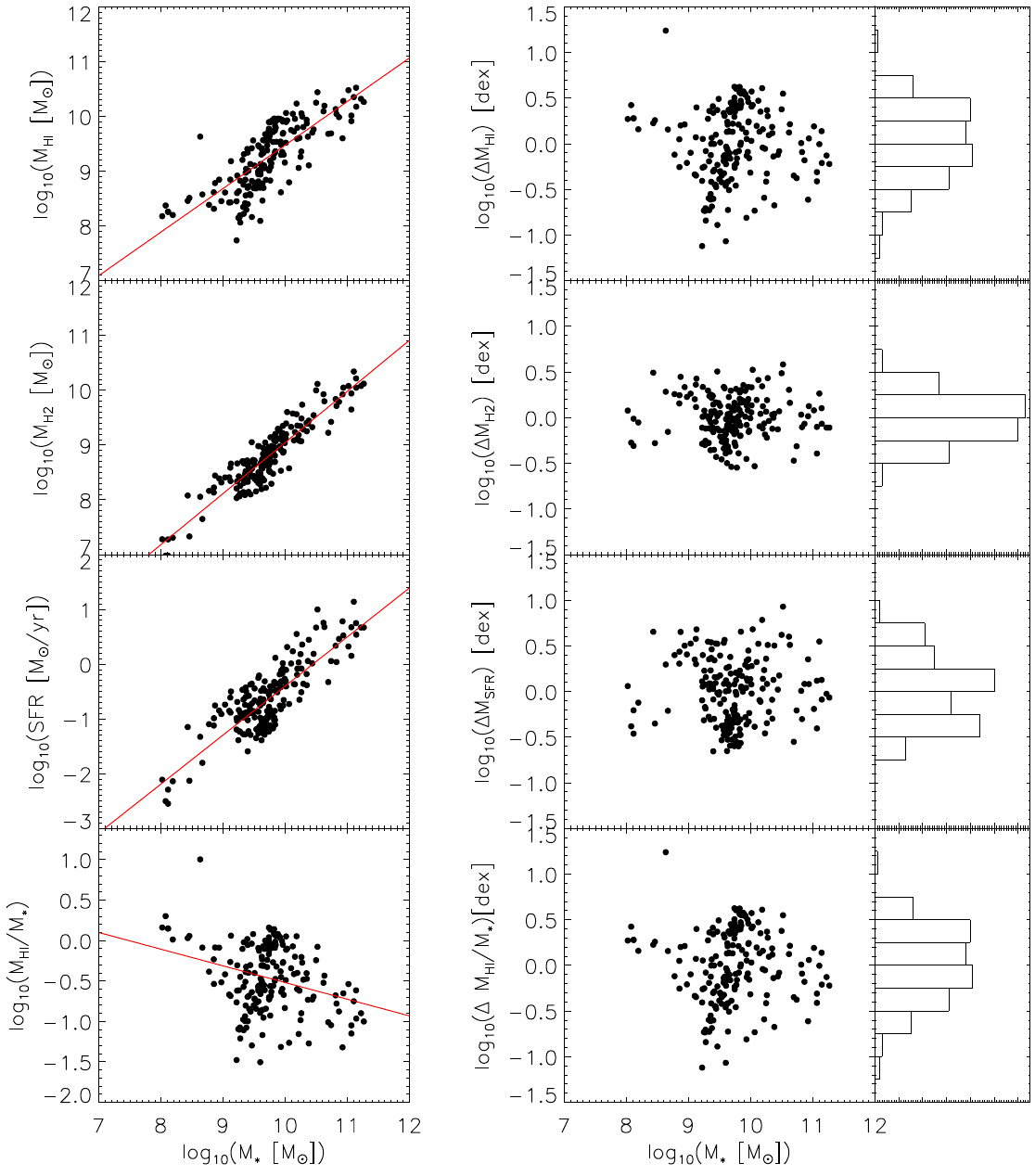}
\caption{Left column: Scaling relations for the 187 galaxies in our \textsc{Simba} sample.  From top to bottom, the relations are \hi\ mass, H$_2$ mass, star formation rate and \hi\ mass fraction as a function of stellar mass.  The red line in each panel is a straight line fit to the data, representing the mean trend with stellar mass.  Middle column: Residuals measured as vertical separations between the data points and the best-fit relation, shown as a function of stellar mass.  Right column: Distribution of the aforementioned residuals.  In Figure~\ref{resid_compare}, we compare the $\Delta j_*$ residuals shown in Figure~\ref{j-M_profiles1} to the residuals shown here.}
\label{props_vs_Mstar}
\end{figure*}

In this section, we search for evidence of links between $j_*$--$M_*$ scatter and various galaxy properties.   Figure~\ref{props_vs_Mstar} shows in its left column several important galaxy scaling relations that are produced by \textsc{Simba}: \hi\ mass ($M_\mathrm{HI}$), H$_\mathrm{2}$ mass ($M_\mathrm{H2}$), star formation rate (SFR) and \hi\ mass fraction ($M_\mathrm{HI}/M_*$ ) as functions of $M_*$ for the 187 galaxies in our sample.  A strong power-law dependence on $M_*$ exists for all of the galaxy properties\footnote{Note that these relations are not generally true when considering all \textsc{Simba} galaxies (e.g., \citealt{Dave_2020}), but only for the sample of star-forming disks we have selected.}.  To each set of points we fit a straight line (in $\log$--$\log$ space) to parameterise the mean trend.  The panels in the middle column of Figure~\ref{props_vs_Mstar} show the residuals calculated by subtracting each set of points from its best-fitting line, as a function of stellar mass.  The panels in the last column of Figure~\ref{props_vs_Mstar}  show the 1-dimensional distributions of the residuals.  The outlying galaxy seen in the top and bottom rows of Figure~\ref{props_vs_Mstar} is the only one in our sample that has not undergone any significant merger activity.  This could explain its very high \hi\ mass.

\begin{figure*}
\centering
\includegraphics[width=1.5\columnwidth]{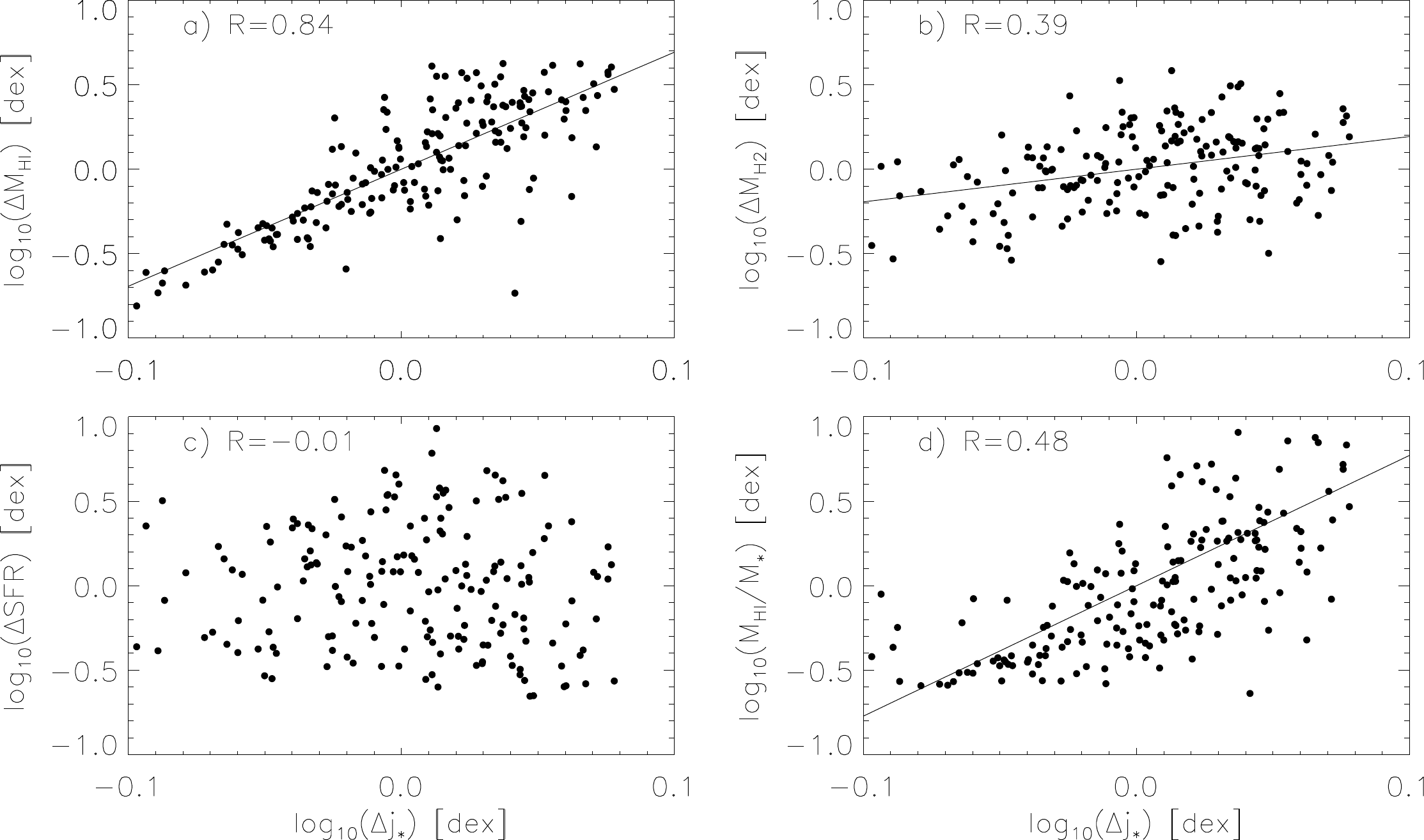}
\caption{Residuals from the $M_\mathrm{HI}$, $M_\mathrm{H2}$, SFR and $M_\mathrm{HI}/M_*$ versus $M_*$ scaling relations (shown in Fig.~ \ref{props_vs_Mstar}) as functions of $j_*$ residuals in the $j_*$--$M_*$ relation.  $j_*$--$M_*$ residuals are clearly strongly related to $M_\mathrm{HI}$ and $M_\mathrm{HI}/M_*$ residuals.  The Pearson correlation coefficient for each data set is shown in the top left of each panel.  The solid lines shown in panels (a), (b) and (d) are fits to the data.  They have slopes of 6.93, 1.93 and 7.72, respectively.}
\label{resid_compare}
\end{figure*}

In Figure~\ref{resid_compare} we compare the $j_*$ residuals from the $j_*$--$M_*$ relation to the residuals from the  scaling relations shown in the left column of Figure~\ref{props_vs_Mstar}.  This allows us to check whether the $j_*$--$M_*$ scatter is correlated with the deviations of galaxies from these important scaling relations.  The results shown in panel (a) of Figure~\ref{resid_compare} clearly indicate that for a given stellar mass, a galaxy's scatter about the $j_*$--$M_*$ relation is highly correlated with its deviation from the mean $M_\mathrm{HI}$--$M_*$ relation.  \hi\ content is a strong driver of $j_*$ residual.  For a given $M_*$, galaxies that have more/less-than-average \hi\ mass also have more/less-than-average specific stellar  angular momentum. The straight line shown in panel (a) is a fit to the data, it has slope $\sim 6.93$.  Thus, a galaxy's $j_*$ deviation of 0.1~dex from the mean $j_*$--$M_*$ typically corresponds to a $\sim 0.7$~dex departure from the mean $M_\mathrm{HI}$--$M_*$ relation.  Panel (b) shows there is a comparatively weak relation between a galaxy's $j_*$ scatter and its H$_2$ content.   Panel (c) suggests there is no dependence of a galaxy's $j_*$ residual on its deviation from the mean SFR--$M_*$ relation.  Finally, panel (d) shows \hi\ mass fraction ($M_\mathrm{HI}/M_*$) to also be strongly linked to a galaxies $j_*$ content, albeit to a lesser extent than \hi\ mass.   

The fact that $j_*$--$M_*$ scatter is linked most strongly to \hi\ content and \hi\ mass fraction is consistent with the results of \citet{Obreschkow_2016}.  Their parameter-free model suggests the neutral atomic mass fractions of disc galaxies to be linearly related to their specific angular momentum content. Galaxies with higher angular momentum content are expected to retain a higher fraction of their neutral atomic has (which is dominated by \hi), hence appearing to have an \hi\ excess relative to galaxies with lower specific angular momentum content.  The filamentary nature of gas accretion could be linked to a galaxy's $j_*$ content.  So-called ``cold flows'' (e.g., \citealt{Keres_2005, Dekel_2009}) are able to transfer gas from the virial radius of a dark matter halo directly to its inner region containing the dominant galaxy.   Gas accreted via this cold mode tends to have higher specific angular momentum than the dark matter halo, some of which is deposited into the central dominant galaxy.  Related to this accretion scenario is the manner in which stellar disks are thought to grow from inside-out.  \citet{Peebles_1969} suggested that when gas disks acquire angular momentum from gravitational torques, is it the low angular momentum gas that most quickly settles and forms stars at inner radii.  Simulations of disk formation (e.g., \citealt{Mo_1998, Somerville_2008}) have shown that galaxies must grow from the inside out in order to reproduce various observed scaling relations. These types of angular momentum acquisition scenarios could be responsible for our observation that for a given stellar mass, \textsc{Simba} galaxies with higher $j_*$ content are those with higher $M_\mathrm{HI}$ content.

\section{Summary}\label{section_summary}
In this work, we have measured the $j$--$M$ relations for the stellar, \hi\ and baryonic mass components for a set of 187 galaxies from the \textsc{Simba} cosmological hydrodynamical simulations.  We compare the measured relations to existing empirical results, and search for systematic trends of residuals in the $j_*$--$M_*$ relation with galaxy properties such as \hi\ mass, $H_2$ mass, star formation rate and \hi\ mass fraction.

For all mass components, we find the specific angular momenta of galaxies to be related via a power law to their masses.  Our measured relations exhibit very little scatter and hold over several orders of magnitude in mass.  The stellar $j$-$M$ relation has a power-law index $\alpha = 0.59 \pm 0.013$, agreeing broadly with the empirical values found in several observational studies, yet with much less scatter, suggesting that measurement uncertainties (related to sample biases and assumptions behind the measurements) are the dominant source of scatter in the observations.  

We compare the residuals of galaxies in the $j_*$--$M_*$ relation to those from other important galaxy scaling relations.  There exists a strong dependence on the \hi\ content of a galaxy.  It is the excess or deficiency of a galaxy's \hi\ mass that is clearly linked to its deviation from the mean $j_*$--$M_*$ relation.  For a given $M_*$, galaxies that have more/less-than-average \hi\ also have more/less-than-average stellar specific angular momentum.  We interpret this as being linked to the \hi\ build-up histories of galaxies, specifically by means of cold mode accretion.

Our finding that the \textsc{Simba} simulations produce tight $j$--$M$ relations that are consistent with empirical results, as well as other important scaling relations that other authors have reported \textsc{Simba} to reproduce, further suggest that \textsc{Simba} will serve as a reliable platform for contextualising the results from ongoing and forthcoming large galaxy surveys carried out with instruments such as MeerKAT, ASKAP and eventually the SKA.

\section{Acknowledgements}
We are sincerely grateful to the anonymous referee for providing very useful and constructive comments that improved the quality of this study.  We thank Robert Thompson for developing \textsc{Caesar}, and the \textsc{yt} team for development and support of \textsc{yt}.  EE acknowledges that this research is supported by the South African Radio Astronomy Observatory, which is a facility of the National Research Foundation, an agency of the Department of Science and Technology.  This work is based on the research project supported wholly/in part by the National Research Foundation of South Africa (grant number 115238).  RD acknowledges support from the Wolfson Research Merit Award program of the U.K. Royal Society.  MG acknowledges support from the Inter-University Institute for Data Intensive Astronomy (IDIA), and by the Australian Government through the Australian Research Council's Discovery Projects funding scheme (DP210102103).

The computing equipment to run \textsc{Simba} was funded by BEIS capital funding via STFC capital grants ST/P002293/1, ST/R002371/1 and ST/S002502/1, Durham University and STFC operations grant ST/R000832/1. DiRAC is part of the National e-Infrastructure.  We acknowledge the use of computing facilities of IDIA for part of this work. IDIA is a partnership of the Universities of Cape Town, of the Western Cape and of Pretoria. We acknowledge the use of the ilifu cloud computing facility --- \url{www.ilifu.ac.za}, a partnership between the University of Cape Town, the University of the Western Cape, the University of Stellenbosch, Sol Plaatje University, the Cape Peninsula University of Technology and the South African Radio Astronomy Observatory. 


\end{document}